\documentclass[BCOR=0mm,DIV=14,fontsize=11pt,titlepage,twocolumn]{scrartcl}
\KOMAoptions{titlepage=false,abstract=true}

\usepackage[utf8]{inputenc}
\usepackage[english]{babel}

\usepackage[autostyle]{csquotes}

\usepackage[letterspace=200]{microtype}

\usepackage{xcolor}

\usepackage[mode=buildnew]{standalone}

\usepackage{tikz}
\usetikzlibrary{shapes}
\usetikzlibrary{decorations.pathreplacing}
\usetikzlibrary{patterns}
\usetikzlibrary{plotmarks}

\usepackage{textcomp}

\usepackage{siunitx}
\DeclareSIUnit{\molar}{\textsc{M}}

\usepackage{xspace}

\usepackage{subfig}

\usepackage{amsmath}

\usepackage{stmaryrd}

\usepackage{listings}
\definecolor{darkgray}{gray}{0.4}
\usepackage{abstract}

\usepackage[colorinlistoftodos]{todonotes}
\usepackage[
  natbib=true,
  style=numeric,
  backref=true,
  backrefstyle=three,
  backend=bibtex,
  firstinits=true,
  maxnames=10,
  minnames=10,
  url=true,
  doi=true,
  eprint=true
]{biblatex}
\definecolor{niceblue}{rgb}{0.122,0.396,0.651}
\newcommand{\pdflinkcolor}{niceblue}

\usepackage{ltxcmds}
\RequirePackage{ifthen}
\makeatletter
\@ifpackageloaded{hyperref}{}{\usepackage[pdftex,%
  colorlinks=true, 
  bookmarks=true, %
  bookmarksnumbered=true, 
  bookmarksopen=true, 
  pdfstartview=FitH, 
  linkcolor=\pdflinkcolor, 
  anchorcolor=\pdflinkcolor, %
  citecolor=\pdflinkcolor, 
  urlcolor=\pdflinkcolor, 
  filecolor=\pdflinkcolor, %
  hyperindex=true, 
  plainpages=false, 
  breaklinks=true, 
  pdfpagelabels, 
  hypertexnames=true, %
]{hyperref}}

\usepackage{cleveref}

\usepackage{doi}

\usepackage[acronym,toc,nomain]{glossaries}
\makeglossaries
\newacronym{LSA}{LSA}{line source approximation}
\newacronym{PNP}{PNP}{Poisson-Nernst-Planck}
\newacronym{ED}{ED}{electrodiffusion}
\newacronym{LFP}{LFP}{local field potential}
\newacronym{DOFs}{DOFs}{degrees of freedom}
\newacronym{AP}{AP}{action potential}
\newacronym{HH}{HH}{Hodgkin-Huxley}
\newacronym{ODE}{ODE}{ordinary differential equation}
\newacronym{PDE}{PDE}{partial differential equation}
\newacronym{FEM}{FEM}{finite element method}
\newacronym{cG}{cG}{continuous Galerkin}
\newacronym{dG}{dG}{discontinuous Galerkin}
\newacronym{GHK}{GHK}{Goldman-Hodgkin-Katz}
\newacronym{PCM}{PCM}{parallel conductance model}
\newacronym{ILU}{ILU}{inexact LU}
\newacronym{BiCGStab}{BiCGStab}{stabilized biconjugate gradient}
\newacronym{AMG}{AMG}{algebraic multigrid}
\newacronym{GMRes}{GMRes}{generalized minimal residual}
\newacronym{TMP}{TMP}{template meta program}
\newacronym[\glslongpluralkey={degrees of freedom}]{DOF}{DOF}{degree of freedom}
\newacronym{ISTL}{ISTL}{iterative solver template library}
\newacronym{EEG}{EEG}{electroencephalography}
\newacronym{ES}{ES}{extracellular space}
\newacronym{VC}{VC}{volume conductor}
\newacronym{HPC}{HPC}{high performance computing}
\newacronym{LS}{LS}{linear solver}
\newacronym{EDL}{EDL}{electric double layer}
\newacronym{CRTP}{CRTP}{curiously recurring template pattern}
\newacronym{IO}{IO}{input/output}
\newacronym{EAP}{EAP}{extracellular action potential}
\newacronym{EN}{EN}{electroneutral model}
\newacronym{EM}{EM}{electron microscopy}
\newacronym{DTI}{DTI}{diffusion tensor imaging}

\usepackage{pods_thesis}

\newcommand{\mycaption}[2][]{\caption[#1]{\textbf{#1}. #2}}

\author{Jurgis Pods \thanks{Interdisciplinary Center for Scientific Computing,
University of Heidelberg, Im Neuenheimer Feld 368, 69120 Heidelberg, Germany. E-mail: \href{mailto:jurgis.pods@iwr.uni-heidelberg.de}{jurgis.pods@iwr.uni-heidelberg.de}}}
\title{A Comparison of Computational Models for the Extracellular Potential of Neurons}

\begin{document}
\twocolumn[
\begin{@twocolumnfalse}
 \maketitle
 \begin{abstract}
   The extracellular space has an ambiguous role in neuroscience. 
   It is present in every physiologically relevant system and often used as a measurement site in experimental recordings, but it has received subordinate attention compared to the intracellular domain.
   In computational modeling, it is often regarded as a passive, homogeneous resistive medium with a constant conductivity, which greatly simplifies the computation of extracellular potentials.
   However, recent studies have shown that local ionic diffusion and capacitive effects of electrically active membranes can have a substantial impact on the extracellular potential.
   These effects can not be described by traditional models, and they have been subject to theoretical and experimental analyses.
   We strive to give an overview over recent progress in modeling the extracellular space with special regard towards the concentration and potential dynamics on different temporal and spatial scales.
   Three models with distinct assumptions and levels of detail are compared both theoretically and by means of numerical simulations: the classical volume conductor (VC) model, which is most frequently used in form of the line source approximation (LSA); the very detailed, but computationally intensive Poisson-Nernst-Planck model of electrodiffusion (PNP); and an intermediate one called the electroneutral model (EN).
   The results clearly show that there is no one model for all applications, as they show significantly different responses especially close to neuronal membranes.
   Finally, we list some common use cases for model simulations and give recommendations on which model to use in each situation.
   \vspace{1cm}
 \end{abstract}
\end{@twocolumnfalse}
]
\saythanks
   
\section{Introduction}
Computational models play an important role for the analysis of complex systems like the brain. 
When applied under the correct assumptions, they allow to obtain results of a system with reduced complexity in order to study the influence of the essential mechanisms.
Consequently, computational models have established as an important tool next to experiments in neuroscience.
A tremendous amount of work has gone into developing models of neurons, pioneered by the work of Hodgkin and Huxley \cite{hodgkin1952quantitative} for the dynamics of membrane currents and extended by Rall \cite{rall1989cable}, who applied \emph{cable theory} to account for the tree-like neuronal morphology.
Others have built upon this work to include complicated geometries, a plethora of different channel types and kinetics, synaptic currents, and more.
These models based on the cable equation, which we here refer to as \emph{\gls{HH}-type} models, are arguably among the most successful models in the natural sciences, demonstrated by the spread of the well-known simulators for these models, NEURON \cite{hines1997neuron} and GENESIS \cite{bower1995book}.

Considering the impressive success of neuron models, it is surprising how little attention the \gls{ES} has received.
Given the fact that more and more experimental recordings are performed extracellularly, one is interested in the process of the generation of extracellular potentials -- and in models that replicate such recordings.
Models producing extracellular potentials will necessarily have to include the extracellular space to some degree.
In \gls{HH}-type models, the \gls{ES} is assumed to be isopotential and most commonly set to a grounding potential of \SI{0}{\volt}.
Such an assumption might be valid when one is interested in the intracellular or membrane potentials only, but it is obviously not useful when regarding extracellular potentials.

Most models for the extracellular potential are based on \gls{VC} theory \cite{plonsey1995volume}, where the \gls{ES} is assumed to be electroneutral (and in most cases also homogeneous), i.e.~any concentration effects by redistribution of ionic charges are neglected.
The relevant parameter for the extracellular medium in these models is the conductivity $\kappa$ (or equivalently, its inverse, the resistivity $\rho$).
Mathematically, the model is obtained by reduction of Maxwell's equations to the electrostatic part, such that the membrane is the only current source contributing to the extracellular potential.
These current sources can be imposed as boundary conditions of a Laplace equation, an elliptic \gls{PDE} which has received a fair amount of theoretical analysis and is considered relatively easy to solve numerically when the conductivity field is not too heterogeneous (see, e.g.~\cite{agudelo2013numerical}).

If the conductivity is furthermore assumed to be homogeneous, an analytical solution commonly referred to as the \gls{LSA} \cite{holt1997critical} can be expressed in cylinder coordinates, where the membrane surface is collapsed to a line source.
This avoids the need for a numerical solution and is computationally tractable, since one only has to compute it at the points of interest.
It has also shown to give quite accurate results at distances larger than about \SI{1}{\micro\metre} from the membrane in an experimental comparison \cite{gold2006origin}.

These \gls{VC}-type models have been refined to represent an inhomogeneous extracellular space \cite{bedard2004modeling} that accounts for effects like frequency filtering \cite{PhysRevE.73.051911,bedard2009macroscopic}.
An interesting technique in this context is the application of inverse methods to these kinds of models, enabling the estimation of current source densities from \gls{LFP} measurements \cite{pettersen2008estimation}.
 
\gls{VC}-type models are based on neglecting any effects of concentrations dynamics on the extracellular potential, which is one central point of criticism.
Charge redistributions by either neural membrane dynamics or other processes regulating the ionic milieu (like buffering or uptake through glial cells and astrocytes) cause concentrations gradients, which induce diffusive currents.
Lately, these concentration effects on the extracellular potential have been recognized.
 In \cite{2015arXiv150707317G}, a new recording technique is suggested to account for the frequency-filtering property of diffusive \gls{ES}, while \cite{2015arXiv150506033H} describes a model including ionic diffusion explicitly in the calculation of the extracellular potential.

The main reason for questioning the assumption of a ``passive'' and electroneutral \gls{ES}, however, is the effect of membrane dynamics.
The membrane can be regarded as an electrochemical capacitor which attracts clouds of ions on both interfaces of the electrolytic solution.
The resulting charge accumulation forms the \emph{Debye layer}, a very thin region around the membrane with steep concentration and potential gradients with a thickness of the order of about \SI{1}{\nano\metre} under physiological conditions.
This layer is affecting the potential in the vicinity of the membrane directly through its electrostatic potential and indirectly, through capacitive currents due to a dynamically changing membrane potential, e.g.~during an \gls{AP}. 
See \cite[chapter 12]{lipowsky1995structure} for a summary of the underlying biophysical theory.

To address these complicating effects on the extracellular potential, we have previously addressed more general models based on the \gls{PNP} system of electrodiffusion, which allows to explicitly model ion concentrations and their dynamics \cite{pods2013electrodiffusion}.
We could show that in contrast to \gls{VC}-type models, the full \gls{PNP} system allows to capture these effects, which have a significant influence on the extracellular potential especially at small membrane distances.

In the same context, a series of publications by \citeauthor{mori2006three} contains detailed analyses of the \gls{PNP} system, resulting in a hierarchy of models with successively reduced complexity, namely the newly developed \emph{electroneutral} model \cite{mori2006three,0901.3914,mori2009numerical}.
Compared to the \gls{PNP} model, the electroneutral model does not require to finely resolve the Debye layer, as its capacitive effects are incorporated implicitly as boundary terms on the membrane.
Note that the model introduced in \cite{2015arXiv150506033H} essentially describes the finite volume solution of the electroneutral model, with the exception that it does not account for Debye layer effects at the membrane boundary layer.
It will therefore not be considered separately in the following.

What all of these \gls{PNP}-type models have in common is their increased computational demand.
Since existence and uniqueness of analytical solutions of the underlying systems of \glspl{PDE} have only been shown for certain special cases \cite{jerome1991finite,biler1994debye,wu2013global,eisenberg2007poisson,schoenke2012,pabst2014}, they have to be solved numerically. 
This requires domain knowledge for the numerical analysis and its (possibly parallelized) solution, which might be one reason that these rather intricate models have not seen a wide distribution.

In this study, we strive to give an overview over recent progress in modeling the extracellular potential of neurons and to compare the models both theoretically and numerically.
We list the assumptions underlying each of the models and the requirements imposed by the respective solution procedures.
This results in a trade-off between accuracy and model complexity.
We mention common use-cases and, ultimately, give recommendations on which model to use (and which not to use) in each of those cases.

\section{Model Theory}
\begin{figure*}
\begin{center}%
\subfloat{%
\centering%
\includegraphics[width=0.59\linewidth]{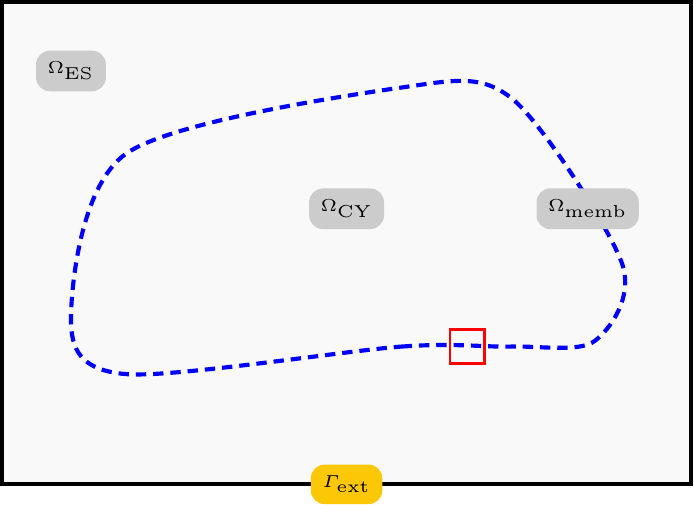}%
}%
\subfloat{%
\centering%
\hspace{0.2cm}%
\raisebox{1.5cm}{%
\includegraphics[width=0.39\linewidth]{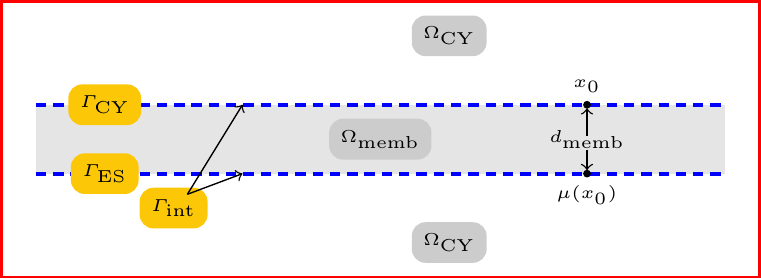}%
}}
\end{center}%
\mycaption[Domain overview and boundary definitions]{%
Figure reproduced with permission from \cite[Fig.~1]{mori2008ephaptic}}%
\label{fig:domain}
\end{figure*}

A schematic view of the computational domain is given in \cref{fig:domain}, consisting of an intracellular domain $\OmegaCytosol$ and the extracellular space $\OmegaExtra$, separated by the membrane interface $\GammaInt$.

The zoom-in shows the membrane subdomain $\OmegaMemb$, delimited by two membrane interfaces.
The internal boundary $\GammaInt = \GammaIntCytosol \bigcup \GammaIntExtra$ therefore consists of two non-connected parts separated by the membrane thickness $\dMemb$.
Each point $x_0 \in \GammaIntCytosol$ on the cytosol-membrane interface is associated with a point $\mu(x_0) \in \GammaIntExtra$ on the opposite membrane-extracellular interface by a map $\mu$.
The values of potential and concentrations evaluated at these points are denoted $\phi^{\text{CY}}(x_0) = \phi(x_0)$, $c_i^{\text{CY}}(x_0) = c_i(x_0)$, $\phi^{\text{ES}}(x_0) = \phi(\mu(x_0))$ and $c_i^{\text{ES}}(x_0) = c_i(\mu(x_0))$ for all $x_0 \in \GammaIntCytosol$.

Here we assume that there are no ion flows present on the membrane domain $\OmegaMemb$, i.e.~we do not explicitly model the microscopic particle flows inside ion channels, but rather resort to the well-established approach of \gls{HH}-type models and represent ion channels by effective conductances and current densities.

This means the Nernst-Planck equations are solved only on the non-connected electrolyte domain $\OmegaElec = \OmegaCytosol \bigcup \OmegaExtra$, and additional Neumann flux conditions are imposed at the membrane interfaces $\GammaInt$ by coupling with the dynamic \gls{HH} channel conductances. 

The mathematical models considered are given in the following.
For the \gls{PNP} model, the Poisson equation is defined on the whole domain. In the case of the \gls{EN} model, another internal boundary condition is imposed for the potential, completely excluding $\OmegaMemb$ from the computational domain.
Each equation is allowed its own partition of the boundaries $\GammaExt$ and $\GammaInt$ into Dirichlet and Neumann conditions, e.g.~the \textbf{N}eumann part of the \textbf{ext}ernal boundary for the \textbf{N}ernst-\textbf{P}lanck equation is denoted $\GammaExtN^{\text{NP}}$.

\subsection{PNP}
We start with the most general model, the \gls{PNP} model consisting of the Nernst-Planck equation
\begin{subequations}
\label{eq:np}
\begin{align}
\frac{\partial c_i}{\partial t}+\nabla\cdot\mathbf{F}_i &= 0 \label{eq:np_main}
\intertext{with the ion flux}
\mathbf{F}_i &= -D_i\left(\nabla c_i + \frac{z_i e}{k_B T} c_i\nabla\phi\right) \label{eq:np_flux}
\intertext{and boundary conditions}
e z_i \mathbf{F}_i \cdot \mathbf{n} &= j_i^{\text{(NP)}} \quad \text{on } \GammaExtN^{\text{(NP)}} \cup \GammaInt^{\text{(NP)}} \label{eq:np_bc_N}\\
c_i &= g_i^{\text{(NP)}} \quad \text{on } \GammaExtD^{\text{(NP)}} \ ,
\label{eq:np_bc_D}
\end{align}
\end{subequations}
where $c_i, \ i=1,\ldots,N$ are the ionic concentrations with units \si{1 \per\cubic\metre} for the $N$ different ion species, $\phi$ is the electric potential with units \si{\volt}, $z_i$ is the valence and $D_i$ the (possibly position-dependent) diffusion coefficient of ion species $i$, $e$ is the elementary charge, $k_B$ the Boltzmann constant, and $T$ is the temperature in \si{\kelvin}.
Together with the Poisson equation for the electric potential
\begin{subequations}
\label{eq:p}
\begin{align}
\nabla\cdot\left(\epsilon\nabla\phi\right) &= -\frac{1}{\epsZero} \left(\rho_0 + \sum_i z_i e c_i \right)
\label{eq:p_main}
\intertext{and boundary conditions}
\epsilon\nabla\phi \cdot \mathbf{n} &= j^{\text{(P)}} \quad \text{on } \GammaExtN^{\text{(P)}} \label{eq:p_bc_N} \\%
\phi &= g^{\text{(P)}} \quad \text{on } \GammaExtD^{\text{(P)}} \ ,\label{eq:p_bc_D}%
\end{align}
\end{subequations}
this constitutes the \gls{PNP} system. 
Here, $\rho_0$ is a fixed background charge density, $\epsilon$ is the relative permittivity and $\epsZero$ the vacuum permittivity.

The boundary conditions at the internal membrane interfaces deserve special attention.
While the Poisson \cref{eq:p_main} can be defined on the whole domain and therefore does not need any additional boundary conditions, the Nernst-Planck \cref{eq:np_main} is only defined on electrolyte subdomains.
The membrane flux condition for species $i$ is
\begin{align}
j_i &= J_i^{\text{memb}}(\mathbf{x}) = J_i^{\text{memb}}(\mu(\mathbf{x})) \nonumber \\
&= \sum_j g_j \frac{1}{e z_i} \left(\membPot + \frac{k_B T}{e z_i} \ln \frac{c_i^{\text{ES}}}{c_i^{\text{CY}}}
\right) \ . \label{eq:Jmemb}
\end{align}
Here, we have have defined the membrane potential $\membPot = \phi^{\text{CY}} - \phi^{\text{ES}}$ and replaced the constant battery $E$ from the \gls{HH} channel current equation by a variable concentration-dependent reversal potential calculated from the Nernst equation.
The voltage-, (possibly) concentration- and time-dependent channel conductances $g_j = g_j(\membPot, c^{\text{CY}}, c^{\text{ES}}, t)$ for each channel type $j$ applying to ion species $i$ can be obtained by coupling with a \gls{HH}-type system. 

\subsection{EN}
The electroneutral model is derived from the \gls{PNP} model by replacing \cref{eq:p} with the electroneutrality condition
\begin{align}
  0 &= \rho_0 + \sum_i z_i e c_i \label{eq:en} \ .
\end{align}
This incorporates the important assumption that the electrolytes are electroneutral at any given time instance, based on the assumption that any charge excess is relaxing quickly towards the electroneutral equilibrium state.
Please note that, although the summed charge density $\sum_i z_i e c_i$ is zero, the individual concentrations are allowed to change in space and time by virtue of \cref{eq:np}.

The constraint \cref{eq:en} can be expressed as a \gls{PDE} by taking the derivative in time and inserting \cref{eq:np}, yielding
\begin{subequations}
\label{eq:en_pde}
\begin{align}
  \nabla\cdot\left(a \nabla\phi + \nabla b\right) &= 0 \label{eq:en_pde_main} \\
\intertext{and boundary conditions}
a\nabla\phi \cdot \mathbf{n} &= j^{\text{(P)}} \quad \text{on } \GammaExtN^{\text{(P)}} \label{eq:en_pde_bc_N} \cup \GammaInt^{\text{(P)}} \\%
\phi &= g^{\text{(P)}} \quad \text{on } \GammaExtD^{\text{(P)}} \ ,\label{eq:en_pde_bc_D}%
\end{align}
\end{subequations}
where
\begin{align}
  a = \sum_i \frac{(z_i e)^2 D_i}{k_B T} c_i && \text{and} && b = \sum_i z_i e D_i c_i \label{eq:cond_theo} \ .
\end{align}
This form is better suited for finite element implementations.

An important modification concerns the membrane boundary conditions.
For the concentrations, we replace \cref{eq:Jmemb} by
\begin{align}
j_i = J_i^{\text{memb}} + \frac{\partial \sigma_i}{\partial t} \ , \label{eq:Jmemb_en}
\end{align}
where the membrane currents $J_i$ are given as above and $\sigma_i$ represent the contributions of each ion species $i$ to the total membrane surface charge density $\sigma = C_m \membPot$ separated by membrane capacitance $C_m$, such that $\sigma = \sum_i \sigma_i$.

By introducing additional state variables $\lambda_i$ on the membrane, the $\sigma_i$ can be defined as
\begin{align}
  \sigma_i &= \lambda_i \sigma \label{eq:en_sigma_i}
\end{align}
where the $\lambda_i$ evolve according to
\begin{align}
  \frac{\partial \lambda_i}{\partial t} &= \frac{\tilde{\lambda}_i - \lambda_i}{\tau} \ , && 
\tilde{\lambda}_i = \frac{z_i^2 c_i}{\sum_k z_k^2 c_k} \ . \label{eq_en_lambda_i}
\end{align}
The last equation describes a relaxation of $\lambda_i$ to the steady-state $\tilde{\lambda}_i$ with a very small time constant of $\tau = \SI{1}{\nano\second}$ which circumvents an instability when choosing $\tau = 0$ ($\Rightarrow \lambda_i = \tilde{\lambda}_i$), see \cite{mori2006three} for details.

As a further intricacy, the potential \cref{eq:en_pde_main} is now defined on the electrolyte domain only, necessitating the addition of internal boundary conditions, given simply as the sum of the membrane currents over each ion species $i$:
\begin{align}
  j^{\text{(P)}} &= \sum_i j_i^{\text{(NP)}} = \sum_i \sigma_i + J_i \nonumber \\
&= C_m \frac{\partial \membPot}{\partial t} + \Iion = \Icap + \Iion \label{Jpot_en} \ .
\end{align}
Here the potential boundary condition incorporates the sum of all membrane currents, the capacitive current $\Icap = C_m \partial \membPot / \partial t$ and the sum of all ion channel currents $\Iion$, demonstrating the direct connection to the cable equation.

\subsection{VC}
The volume conductor approach removes the concentration variables from the system by assuming they are fixed in space and time, i.e.~$\partial c_i / \partial t = 0$ and $\nabla c_i = 0 \ \forall i=1,\ldots,N$.
Inserting this into the \gls{EN} equations removes \cref{eq:np} entirely and \cref{eq:en_pde} reduces to
\begin{align}
  \nabla\cdot\left(\kappa \nabla\phi\right) &= 0 \label{eq:vc} \ ,
\end{align}
where $b=0$ cancels out and $\kappa = a(\mathbf{x},t)$ as given in \cref{eq:cond_theo} is the \emph{conductivity} field.

Note that this relation allows for the calculation of the conductivity at any point in space only from (known) physical parameters and the concentrations at this point.
An equivalent expression is used in \cite{2015arXiv150506033H}.
We were not able to find such a relation in the classical biophysical literature.
It could prove useful in practice, considering e.g.~new imaging techniques like \gls{DTI}, which together with the (known or estimated) electrolyte concentrations enables to easily calculate the conductivity field for a \gls{VC}-type model.

Another notable fact is that the extracellular space is completely ``passive'' in this model.
Since it is derived as a special case of the \gls{EN} system, the electroneutrality condition is inherent.
Additionally, any concentration dynamics have been removed and lumped into a single (but possibly position-dependent) conductivity parameter $\kappa$.
If $\kappa$ is scalar and constant in space, the well-known \gls{LSA} can be used as an analytical solution, which makes it a very convenient model due to the greatly simplified solution procedure.
In any case, the system dynamics are completely determined by the membrane current sources given by \cref{Jpot_en}, which takes the familiar form from the well-known cable equation.

\subsection{Model Hierarchy and Extracellular Regimes}
Apparently, the three models form a hierarchy, as each one is a special case of the previous one, obtained by adding an additional assumption.
This makes the specialized models simpler and in general also easier to solve, but it also restricts their applicability, as the required assumptions do not hold in general.
The reason for this is the presence of cell membranes and their induced Debye layer effects mentioned above.
A detailed dimensional analysis of the considered models can be found in \cite{mori2006three}.
In the following, we reproduce the main result.

The analysis yields three characteristic length scales, which allows to roughly partition the \gls{ES} into three regimes, defined by the set of points with a minimum membrane distance $d$ in a given range.
\begin{itemize}
 \item The Debye layer is characterized by a very small membrane distance of the order of the Debye length, $\bigO{\dDebye}$.
In the following, we define the Debye layer range by the range $d \in [0,\, 10 \dDebye]$. The electroneutrality condition \cref{eq:en} does not hold in this regime due to the existence of steep concentration and potential gradients.
 \item Then there is the bulk solution, where the electrolyte is in electroneutral equilibrium state, for $d > \dBulk$, with $\dBulk \in \bigO{\sqrt{\dDebye}}$.
In the following numerical evaluation, $\dBulk \approx \SI{5}{\micro\metre}$ is a good approximation, but this depends on the chosen setup.
\item Between these two regimes resides the diffusion layer with $d \in (10 \dDebye,\, \dBulk)$, inside which concentrations change in place and time in response to the Debye layer dynamics, whereas the total charge density at each point still sums up to zero, following the electroneutrality condition \cref{eq:en}.
\end{itemize}

The analysis shows that the \gls{PNP} is valid in all three \gls{ES} regimes, the \gls{EN} model is valid in both diffusion layer and bulk solution, and \gls{VC}-type models are valid within the bulk solution only.

\section{Numerical Methods}
Both \gls{PNP} and \gls{EN} models represent coupled systems of \glspl{PDE}, for which no analytical solutions are known, although electrodiffusion systems have received quite some attention in the fields of semiconductor and biomolecule analysis \cite{Lu20106979}.

The numerical code is implemented using the \Dune framework \cite{dunegridpaperI:08,dunegridpaperII:08} consisting of the core modules and following additional modules: 
\begin{itemize}
  \item \lstinline!dune-multidomaingrid! \cite{dune-multidomaingrid,Muething:13193} contains a grid with an arbitrary number of user-defined subdomains;
  \item \lstinline!dune-pdelab! \cite{pdelabalgoritmy} provides a generic interface for the definition of ``local operators'' (containing the weak forms of \glspl{PDE}) and their solution by a range of built-in numerical schemes; and finally
  \item \lstinline!dune-multidomaingrid! \cite{Muething:12887}, which is an add-on to \lstinline!dune-pdelab! providing the multi-physics functionalities for solving different equations on subdomains of a grid from \lstinline!dune-multidomaingrid!.
\end{itemize}
 
The numerical algorithm described in the following has been open-sourced and made publicly available on Github for future reference \cite{pods-dune-ax1}.

We use Q1 Finite Elements to discretize the equations in space and an Implicit Euler method for time-stepping.
There are several options of how to deal with the \gls{PDE} system.
One option is to use an \emph{operator-splitting} and solve the concentration and potential equations alternately until convergence in each timestep.
A detailed numerical scheme using this approach for the \gls{EN} system has been developed in \cite{mori2009numerical}.
It has the nice feature that each of the resulting systems is linear and can be solved directly by application of a linear solver.
However, it introduces a splitting error of the order $\BigO{\dt}$ and might severely restrict the maximum usable time step size $\dt$ to ensure a stable method, which was reported for the \gls{PNP} system in \cite{pods2013electrodiffusion}.
It can be attributed to the need of resolving the Debye layer dynamics, which happen on very small spatial and temporal scales compared to the membrane potential dynamics.
This problem can be circumvented by solving the complete \gls{PDE} system in one go using Newton's method.
While this requires solving a numerically harder problem, it handles the nonlinear coupling between concentration and potential equations directly, which presents the main source of numerical instabilities, and consequently allows for a larger time step size $\dt$.

A further complication is given by the question of how to handle the additional system of \glspl{ODE} present in the dynamic \gls{HH} membrane boundary conditions.
Again, we have the option to split the calculation of the membrane fluxes from the solution of the main system (explicit calculation) or to include it implicitly in a fully-coupled approach.
This choice does not have such a strong impact on the numerical stability, but on another matter. 
In the \gls{EN} model, when using pure Neumann boundary conditions for the potential in at least one subdomain, the solution is only determined up to a constant.
This problem can most easily be solved by an implicit handling of membrane fluxes.

\section{Results}
\subsection{Model Problem}
After deducing the models and solution methods, we are now ready to compare the computed results with each other.
In the following, we will consider a single axon embedded in an extracellular bath as the simulation setup for a simple reason: both \gls{PNP} and electroneutral model have not been implemented in a full 3D setup yet due to the large computational demands and the problem of obtaining a computational grid for a complex extracellular geometry.

Approximating the axon as a cylinder yields a rather simple geometry in cylinder coordinates, for which implementations exist for all of the regarded models.
This enables us to simulate the \gls{EAP} of a single axon, a setup that already provides sufficient complexity to demonstrate the differences of model responses.

Furthermore, the homogeneous \gls{ES} with a constant scalar conductivity $\kappa$ allows to use the computationally advantageous analytical \gls{LSA} for the \gls{VC} solution.
An evaluation comparing analytical \gls{LSA} and numerical \gls{VC} solutions confirming the equivalence in this case can be found in \cite[chapter 5]{pods-phdthesis}.

We used the value of $\kappa = \SI{1.39}{\siemens\per\metre}$ for the \gls{LSA} model, which was originally fitted to calibrate it to the \gls{ES} model in the bulk solution for given extracellular concentrations of $c_{\naIon} = \SI{100}{\milli\molar}$, $c_{\kIon} = \SI{4}{\milli\molar}$, and $c_{\clIon} = \SI{104}{\milli\molar}$ \cite{pods2013electrodiffusion}.
Moreover, it matches remarkably well with the theoretical value as calculated by evaluating the expression for $a$ in \cref{eq:cond_theo}.
Detailed explanation on the calculation of extracellular conductivity and reference values from the literature can be found in \cite[chapter 5.3]{pods-phdthesis} and \cite{2015arXiv150506033H}.
The above choice of concentrations results in an extracellular Debye length of about $\SI{0.9}{\nano\metre}$.

We chose the fully-coupled approach for both \gls{PNP} and \gls{EN} models. 
For optimal comparability, we carried out a \gls{PNP} simulation with explicit membrane flux handling and used the obtained membrane fluxes as boundary conditions for the subsequent \gls{EN} and \gls{LSA} simulations.
This ensures that all three models use exactly the same membrane current sources to calculate the extracellular potential, eliminating any inconsistencies due to numerical errors.

As carried out earlier, the \gls{ES} can be roughly divided into three partitions, depending on the membrane distance $d$: the Debye layer, the diffusion layer (or nearfield), and the bulk solution (or farfield).
We will compare the extracellular potential separately for these three regions to illustrate the differences between considered models.

\subsection{Bulk solution}
\begin{figure*}%
\centering%
\subfloat[]{\includegraphics[width=0.33\textwidth]%
{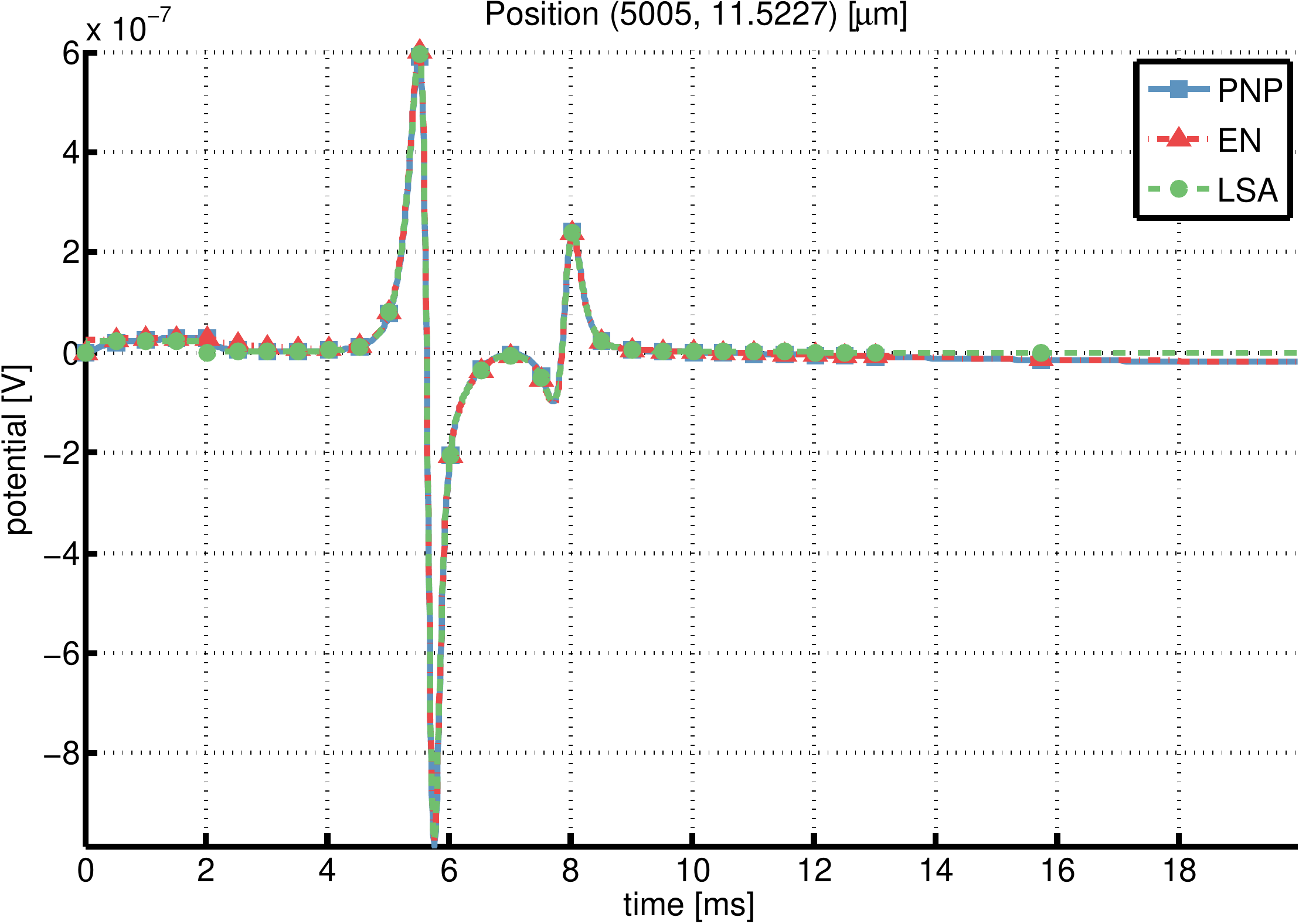}\label{fig:bulk-pnp_mori_lsa1}}%
\subfloat[]{\includegraphics[width=0.33\textwidth]%
{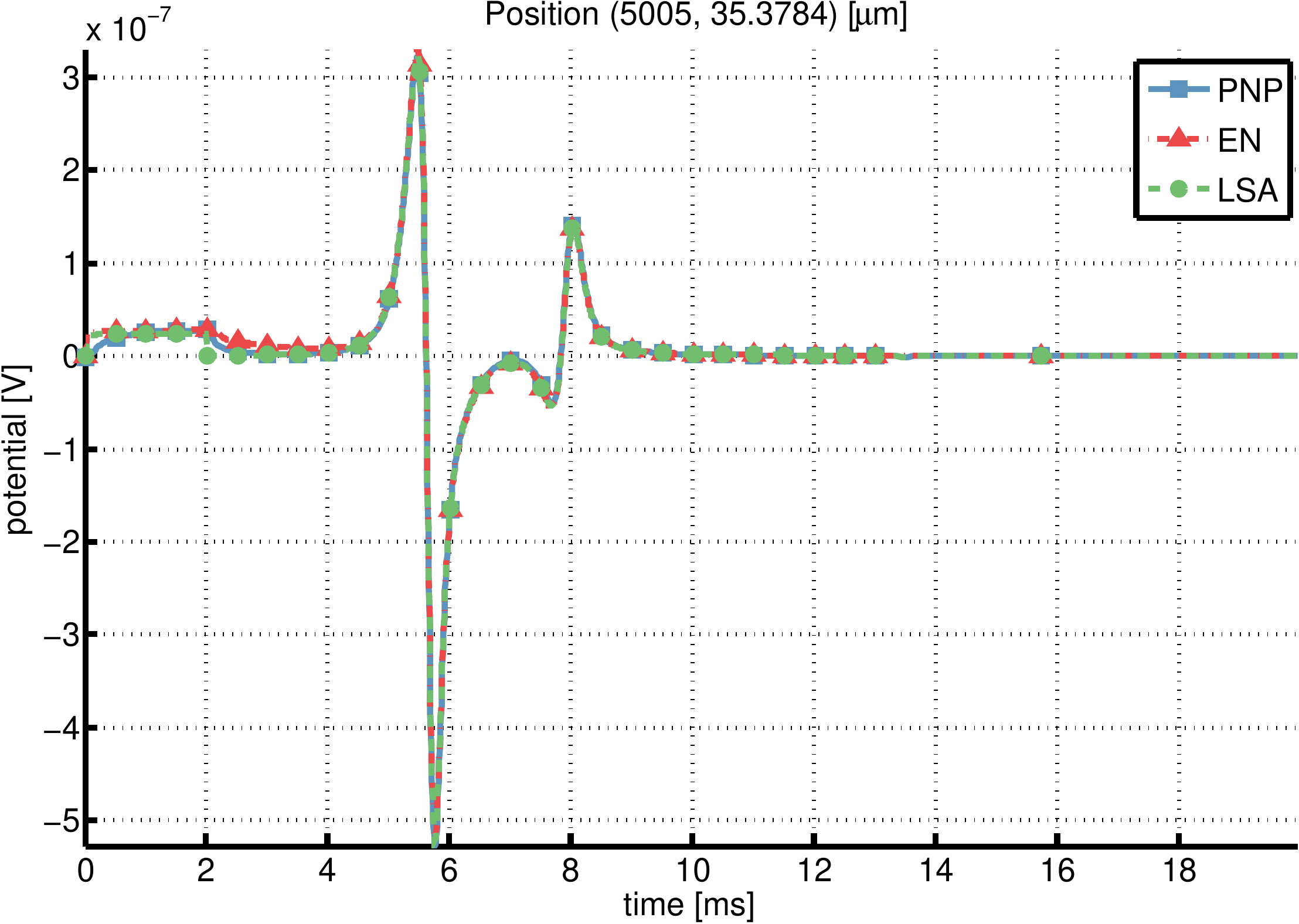}\label{fig:bulk-pnp_mori_lsa2}}%
\subfloat[]{\includegraphics[width=0.33\textwidth]%
{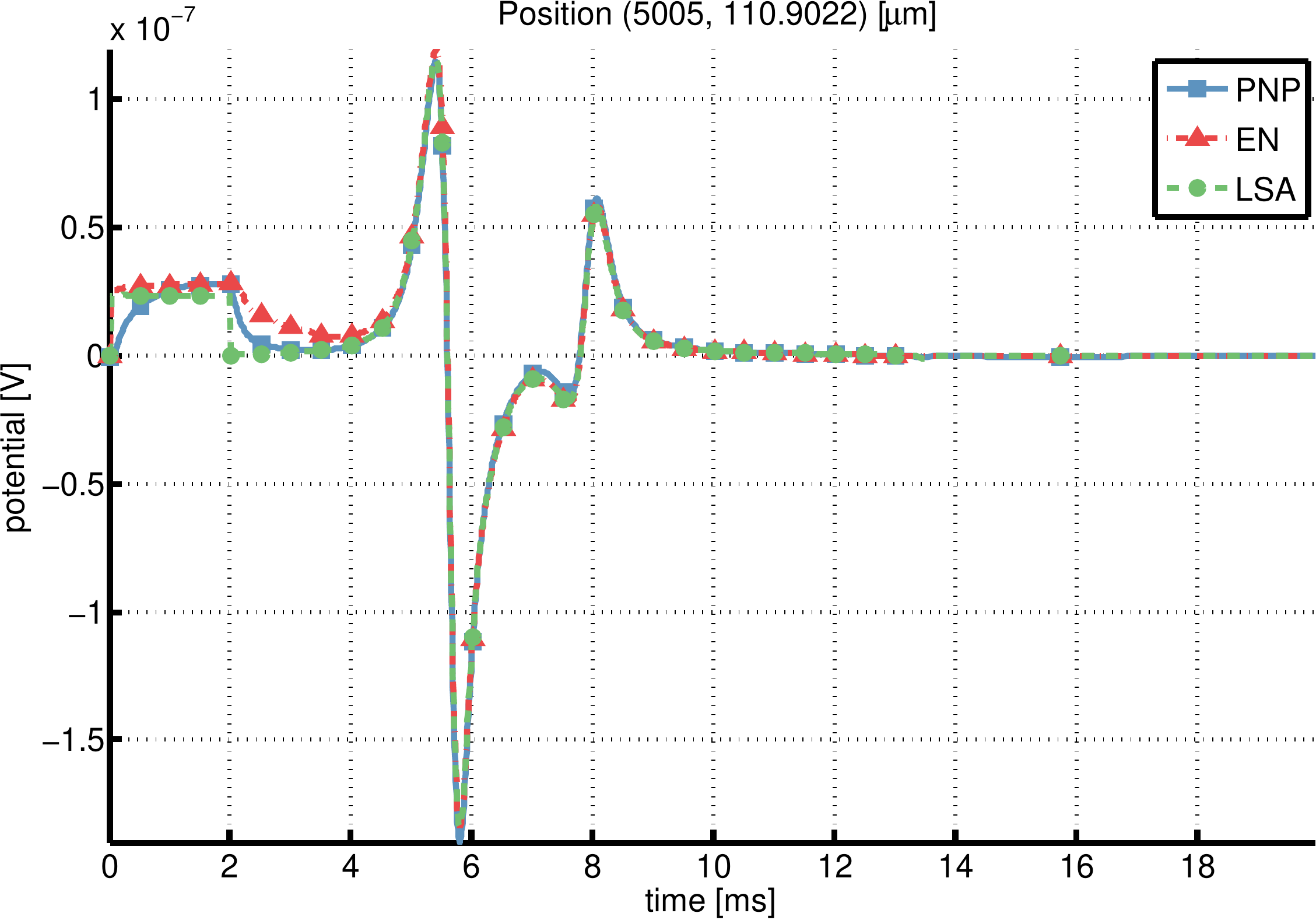}\label{fig:bulk-pnp_mori_lsa3}}%
\mycaption[Comparison of PNP, EN and LSA solutions in bulk solution]{%
}%
\label{fig:bulk-pnp_mori_lsa}%
\end{figure*}

\Cref{fig:bulk-pnp_mori_lsa} shows that at comparably large membrane distances, all three models yield the same potential in response to the axonal \gls{AP}.
This is an important result, as it demonstrates that all the models converge to the same solution in the electroneutral bulk solution and serves as a validation of the numerical implementations of both \gls{PNP} and \gls{EN} models.
The \gls{EAP} is triphasic, as it consists of two major components, which are proportional to the respective membrane currents: the capacitive component, which is responsible for the first peak of the \gls{EAP}, and the ionic component, which constitute the following trough ($\naIon$ inflow) and second peak ($\kIon$ outflow). A detailed analysis can be found in \cite{holt1997critical}.

\subsection{Diffusion layer}
\begin{figure*}%
\centering%
\subfloat[]{\includegraphics[width=0.33\textwidth]%
{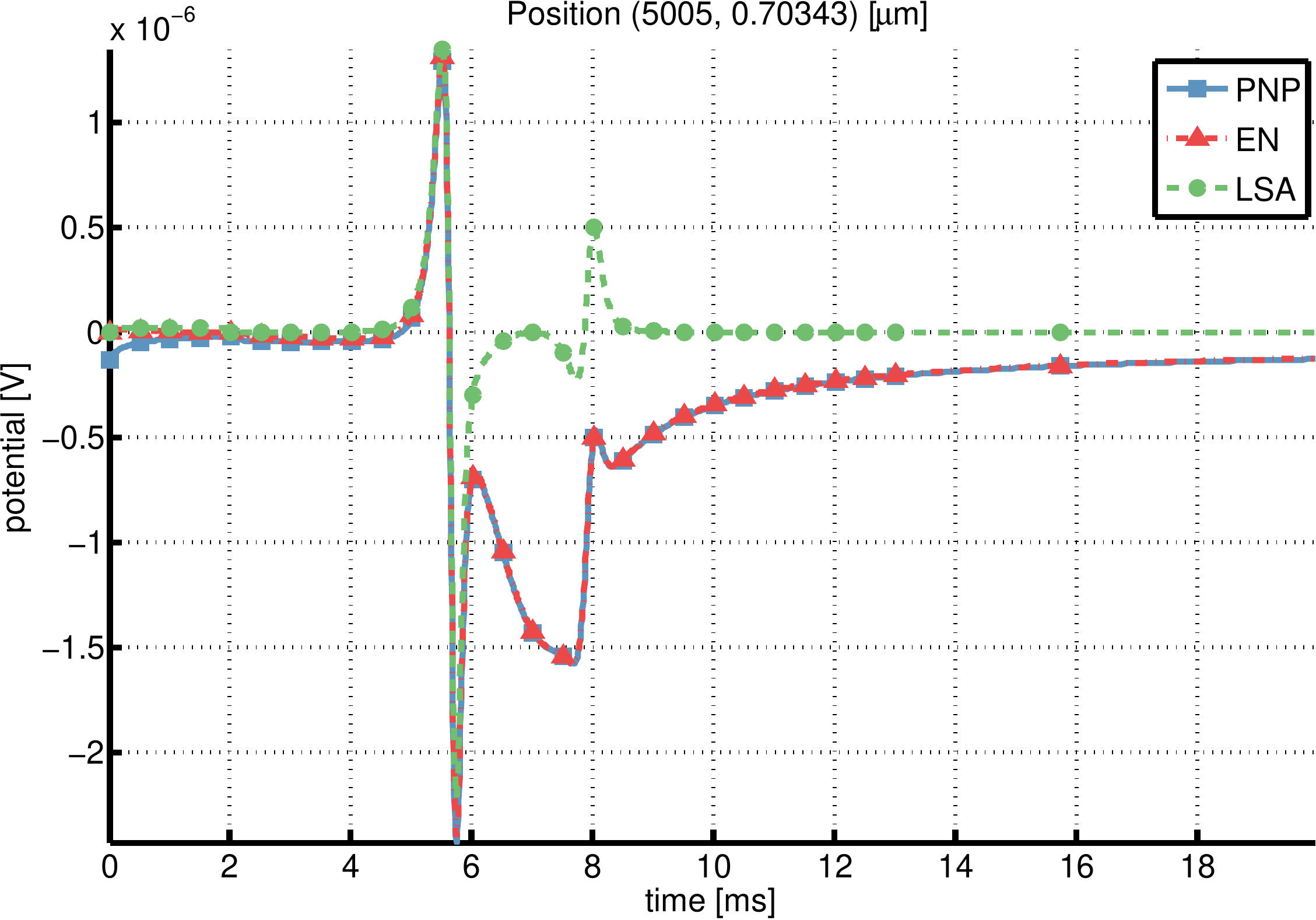}\label{fig:diffusion-pnp_mori_lsa1}}%
\subfloat[]{\includegraphics[width=0.33\textwidth]%
{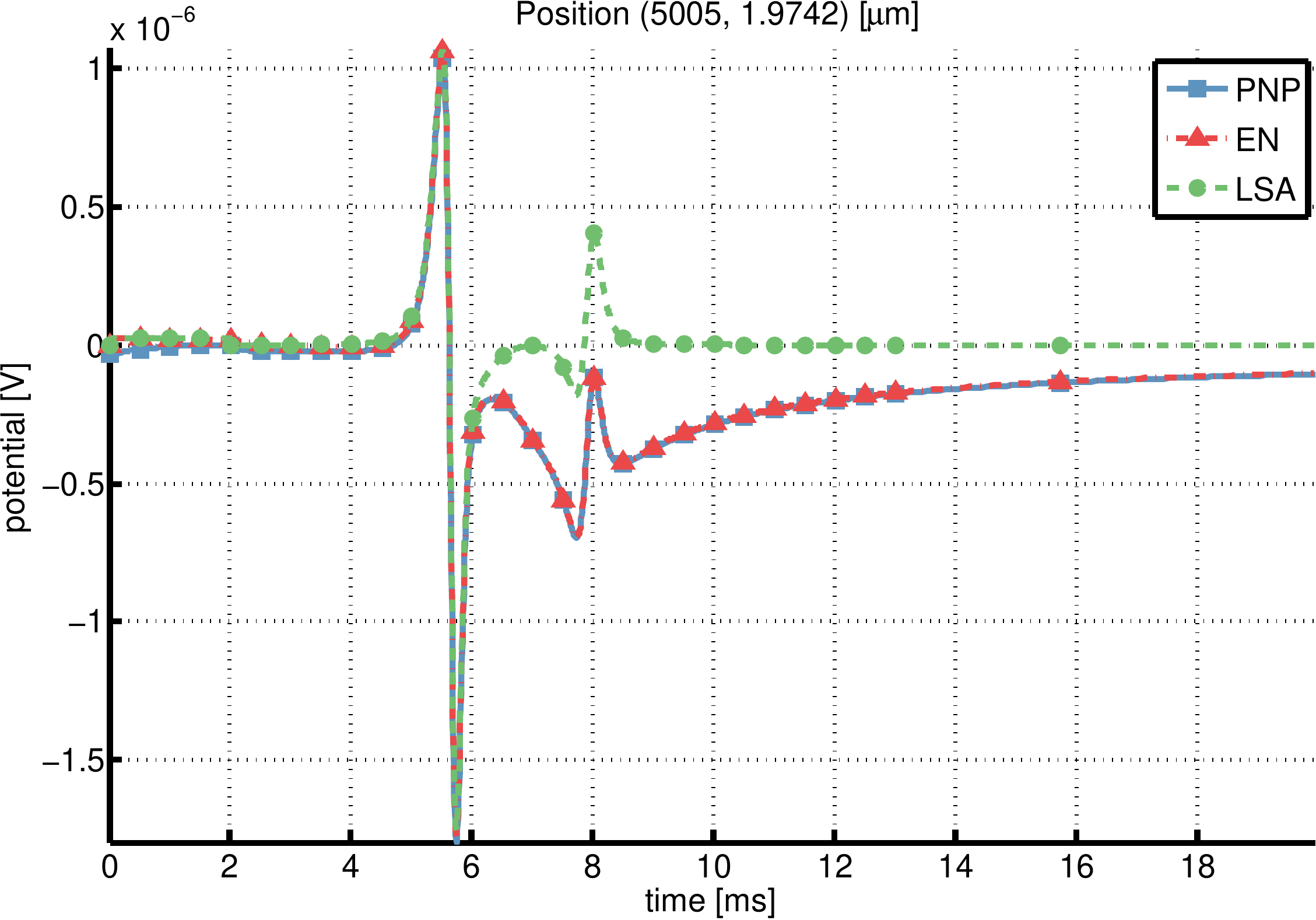}\label{fig:diffusion-pnp_mori_lsa2}}%
\subfloat[]{\includegraphics[width=0.33\textwidth]%
{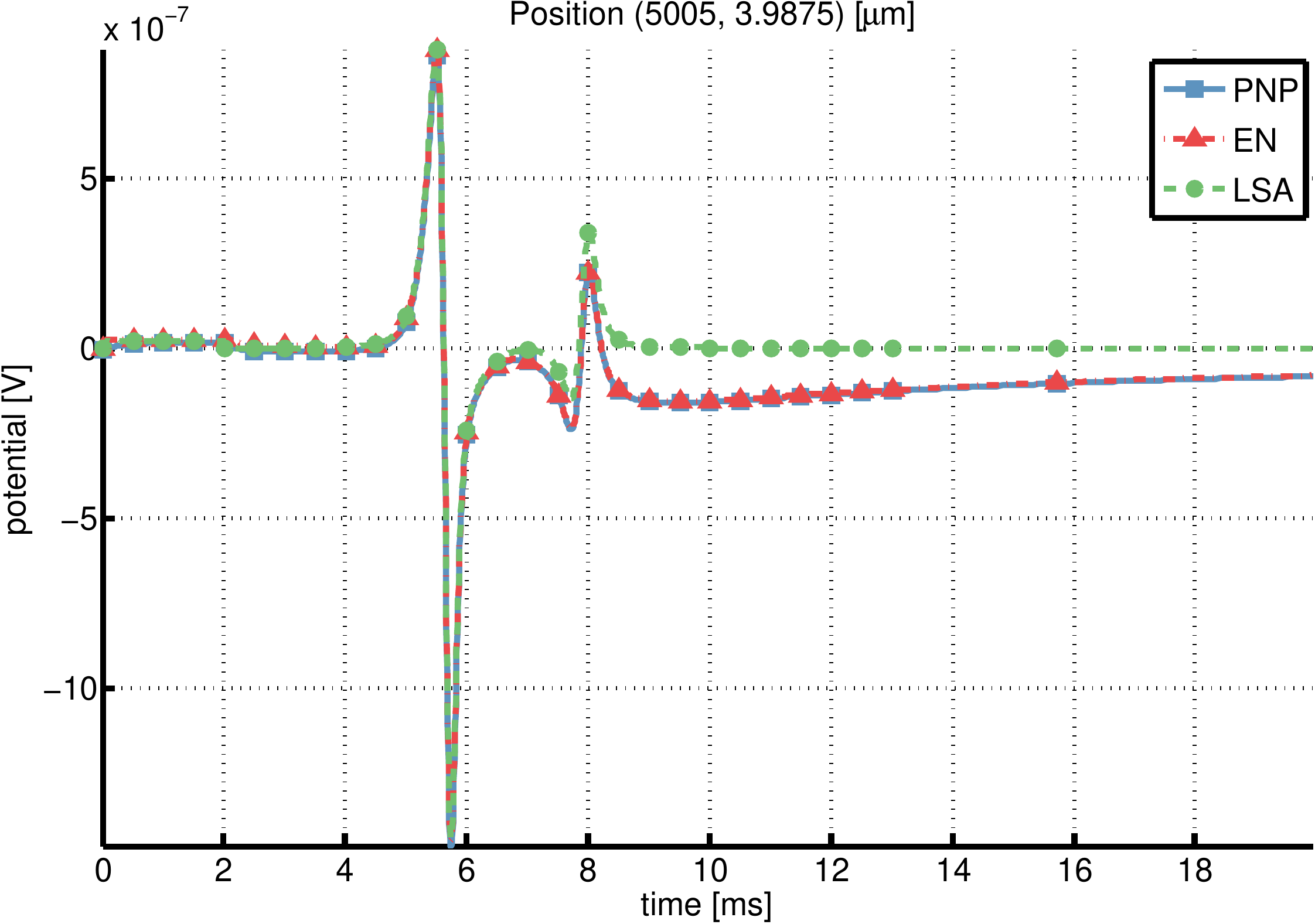}\label{fig:diffusion-pnp_mori_lsa3}}%
\mycaption[Comparison of PNP, EN and LSA solutions in diffusion layer]{%
}%
\label{fig:diffusion-pnp_mori_lsa}%
\end{figure*}

We now turn to the diffusion layer in \cref{fig:diffusion-pnp_mori_lsa}, which shows deviations between \gls{LSA} and the other two models. 
This was to be expected considering our previous theoretical analysis.
In this region, the potential is influenced by ionic redistributions caused by the membrane activity.
Only \gls{PNP} and \gls{EN} models are able to correctly represent these effects.

\subsection{Debye layer}
\begin{figure*}%
\centering%
\subfloat[]{\includegraphics[width=0.33\textwidth]%
{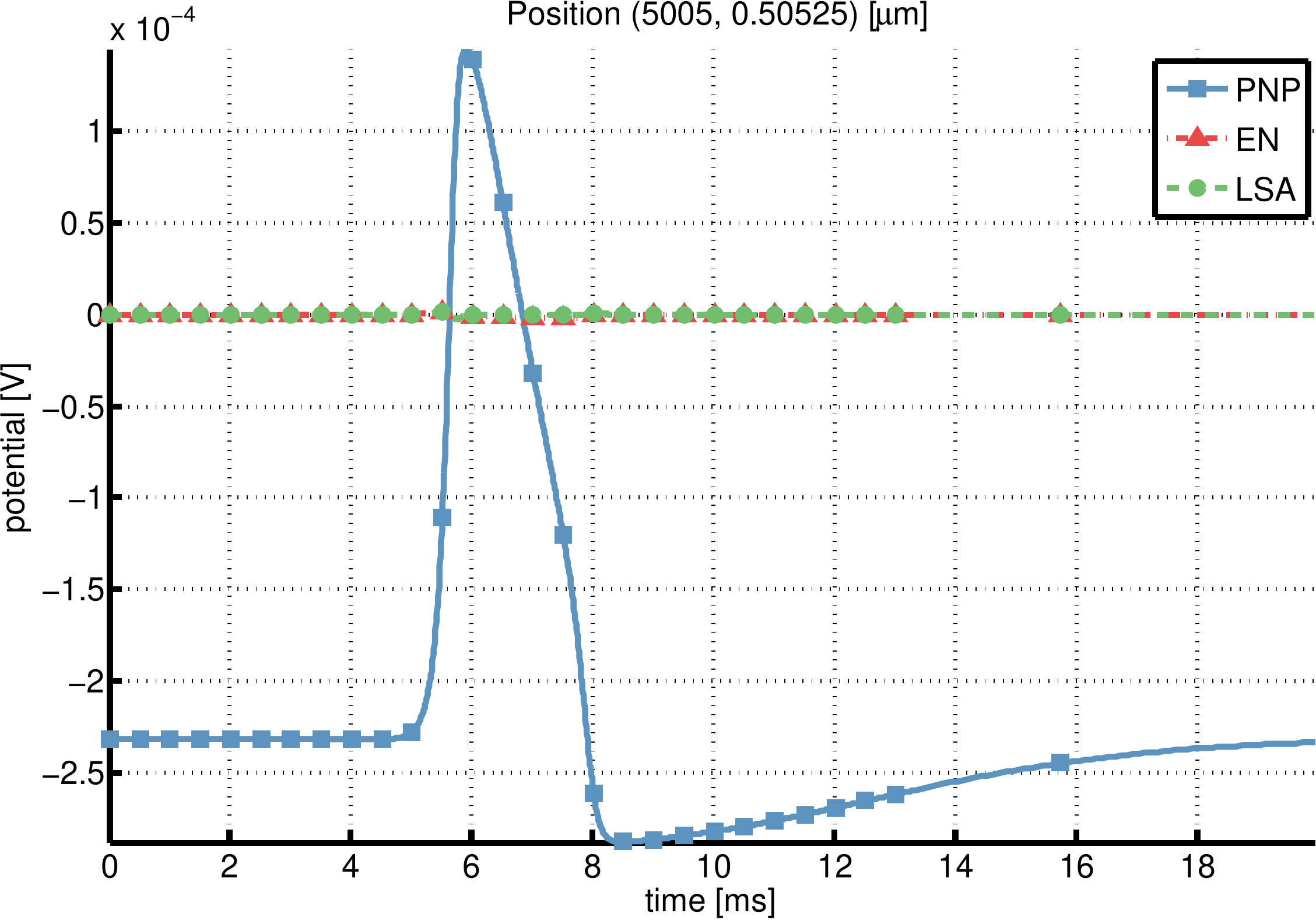}\label{fig:debye-pnp_mori_lsa1}}%
\subfloat[]{\includegraphics[width=0.33\textwidth]%
{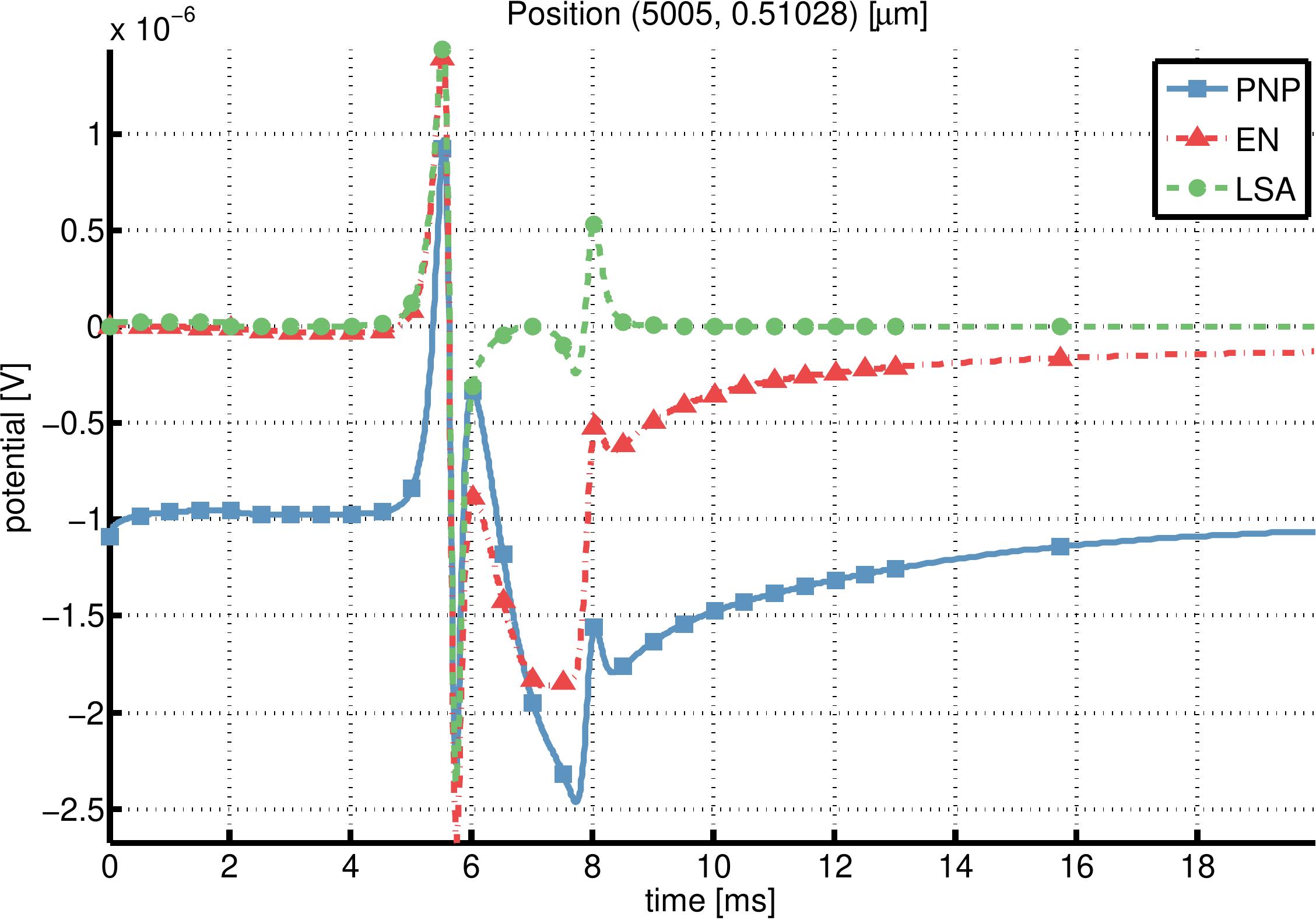}\label{fig:debye-pnp_mori_lsa2}}%
\subfloat[]{\includegraphics[width=0.33\textwidth]%
{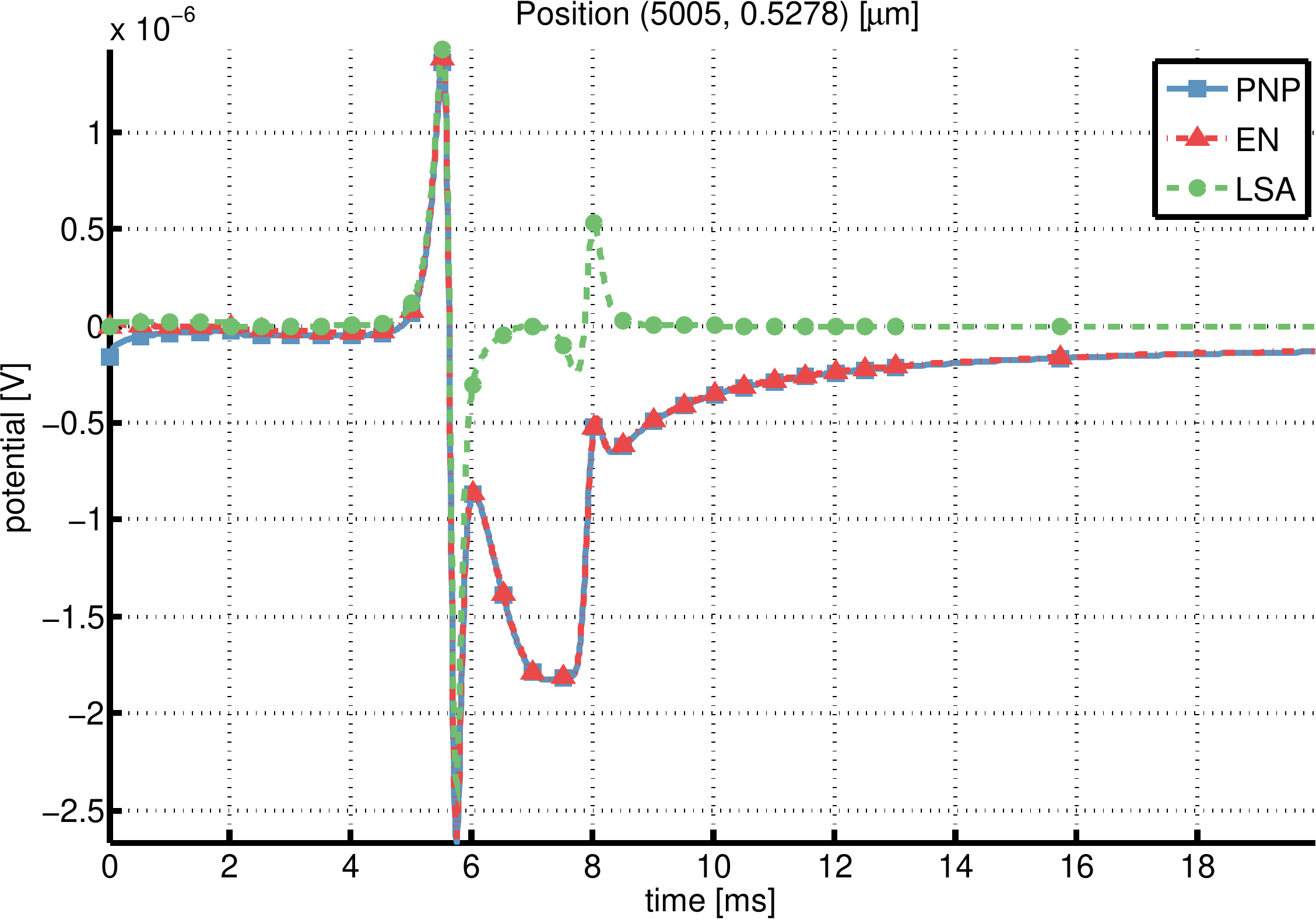}\label{fig:debye-pnp_mori_lsa3}}%
\mycaption[Comparison of PNP, EN and LSA solutions in Debye layer]{%
}%
\label{fig:debye-pnp_mori_lsa}%
\end{figure*}

Finally, we look at the Debye layer very close to the membrane in \cref{fig:debye-pnp_mori_lsa}.
In this region, only the \gls{PNP} model is able to reproduce the ionic gradients and their large influence on the extracellular potential. 
Naturally, these effects are not captured by either \gls{EN} or \gls{LSA}.
In contrast to the previous comparisons, we here simulated the \gls{EN} model with the same grid as the \gls{PNP} model, i.e.~finely resolving the Debye layer, in order to evaluate the potential at the same coordinates.
Note that such a fine resolution is normally not required for the \gls{EN} model.

We see that even with the same spatial resolution, the \gls{EN} model does not reproduce the correct Debye layer potential, as the electroneutrality assumption does not hold in the membrane vicinity \footnote{In \cite{mori2009numerical}, a postprocessing strategy is discussed, which can be applied to correct the \gls{EN} solution with respect to the Debye layer after each time step, which was disregarded for the context of this work.}.
However, it converges quickly to the \gls{PNP} solution and it yields valid results already at a minimum distance of about \SI{20}{\nano\metre}, just outside of the Debye layer, where \gls{PNP} and \gls{EN} solutions coincide.

\section{Discussion}
Recent studies show that the extracellular space should not be regarded as a purely passive Ohmic medium, as in classical volume conductor theory. 
The extracellular modeling community has already recognized the limitations of the established models and recently proposed enhancements to theoretical \cite{2015arXiv150506033H} and experimental methods \cite{2015arXiv150707317G}.
In this work, we have given an overview over current modeling efforts to include influences of both capacitive membrane dynamics and extracellular diffusion on the extracellular potential.

The numerical results confirm what has been found theoretically.
The extracellular domain can be roughly divided into three partitions: bulk solution, diffusion layer and Debye layer.
All three models are valid in the bulk solution.
Only the models that allow for dynamic concentrations yield valid results in the diffusion layer.
This should be emphasized particularly, respecting the fact that the \gls{LSA} model today is widely used without further consideration if it is a valid approximation.

As expected, only the \gls{PNP} model is appropriate for quantitatively calculating the Debye layer potential.
We note that the theoretical hierarchy of models is represented directly by the subsets of the extracellular regime in which each model yields valid results.

As mentioned before, the accuracy of a model does not come without a cost.
While the \gls{LSA} can be implemented easily and solved very fast, both \gls{PNP} and \gls{EN} have to be solved numerically on a computational grid and therefore require expert knowledge and significant simulation times.
Detailed comparisons of the discussed models in terms of computation time and efficiency with respect to the linear and nonlinear solvers were disregarded for the sake of this study.

There is a clear trade-off between accuracy and effort, both computational and implementation-wise, and it is important to know which model should be used in which situation.
The following listing strives to give advice on which model to use depending on the particular case.

\begin{itemize}
  \item \emph{\gls{LFP}}: If one is interested in grand average potentials like \glspl{LFP}, one is commonly dealing with spatial scales of hundreds of micrometers and millimeters.
   For this use-case, the established method of using a cable equation model for the intracellular potential and \gls{LSA} to compute the extracellular response represents the best choice, especially when the potential is only needed at a few distinct measurement points.
   The conductivity parameter should be corrected for the fraction of extra- to intracellular space and tortuosity (hindrances like other cell membranes), see \cite{sykova2008diffusion} for an in-depth exposition.
  
  However, if the extracellular domain is highly inhomogeneous, one should resort to the more flexible \gls{VC} model and explicitly include the spatial conductivity distribution into the model, especially when conductivity-dependent effects like frequency-filtering are considered \cite{bedard2009macroscopic}.
  \item \emph{\gls{EAP}}: For membrane distances in the low and sub-micrometer range, the inclusion of concentration effects is mandatory. 
The \gls{EN} model provides the best trade-off between accuracy and cost. 
It is applicable to a wide range of physiological situations, including the quantitative calculation of \glspl{EAP} and single unit recordings.
  \item \emph{Juxtacellular recordings}: When comparing with juxtacellular recordings, Debye layer effects become the dominating contributions to the extracellular potential and therefore require the full \gls{PNP} model.
  \item \emph{Ephaptic potentials}: In brain areas with restricted conductivity like the tightly packed hippocampus CA1 region, \emph{ephaptic effects} can play an important role and result in unusually large extracellular potentials that may even induce action potentials in neighboring cells, cf.~\cite{jefferys1995nonsynaptic,barr1992electrophysiological}.
  
It has been shown that both \gls{PNP} and \gls{EN} approaches allow to model various ephaptic phenomena that have been experimentally observed, but could previously not be reproduced in models \cite{mori2008ephaptic}\cite[chapter 7]{pods-phdthesis}.
  \item \emph{Complex \gls{ES}}: In any case, when a complex extracellular geometry is considered explicitly, one of \gls{EN} or \gls{PNP} has to be used.
  The reason is the small average membrane distance in the brain, which is estimated to be \SIrange{38}{64}{\nano\metre} \cite{sykova2008diffusion}.
  Consequently, every point in the \gls{ES} will have a very small membrane distance and lie either within Debye or diffusion layer, and \gls{LSA} can not be used to calculate valid results.
  
  This should be taken into account for such models based on high-resolution \gls{EM} reconstructions \cite{denk2004serial}, which have become available recently.
  For these geometries, the \gls{EN} model appears promising, since the mesh generation in 3D is much easier when the Debye layer does not have to be resolved.
\end{itemize}

While \gls{VC}/\gls{LSA} are widely used, further research is needed to actually simulate full 3D models with realistic geometries in reasonable time.
However, one should refrain from simply using the established models for a given problem and carefully consider if they are applicable to the particular situation, as our results show that the validity of the calculated potential critically depends on the membrane distance.

Even if one it not interested in potentials very close to the cell, complicating effects like anisotropic conductivity, frequency filtering or ephaptic potentials may shape the extracellular potential to such a degree that classical models are rendered inapplicable.
Further work is required to enhance the novel models and to meticulously compare them with experimental recordings in different extracellular constellations.

\section*{Acknowledgments}
The author cordially thanks Yoichiro Mori for helpful discussions regarding the implementation of his model and valuable comments on the manuscript.

This work was funded through a grant from the German Ministry of Education and Research 
(BMBF, 01GQ1003A).

\printbibliography[heading=bibintoc]
\newpage

\printglossary

\end{document}